\documentclass[12pt]{article}\usepackage{amsmath,amsfonts}

\newcommand{\bse}{\begin{subequations}}
\newcommand{\ese}{\end{subequations}}
\newcommand{\be}{\begin{equation}}
\newcommand{\ee}{\end{equation}}
\newcommand{\bea}{\begin{eqnarray}}
\newcommand{\eea}{\end{eqnarray}}
\newcommand{\ba}{\begin{array}}
\newcommand{\ea}{\end{array}}

\newcommand{\bc}{{\bar c}}
\newcommand{\bZ}{{\bar Z}}
\newcommand{\bb}{{\bar b}}
\newcommand{\bd}{{\bar d}}

\def\3dNe{$3d,\ {\cal N}=8\ $}

\def\sun2{$su(N)\times su(N)$}

\def\susy{supersymmetry}

\def\id{1\!\!1}
\makeatother
\def\by{\times}
\def\A{{\cal A}}
\begin{document}
\baselineskip 18pt%

\begin{titlepage}
\vspace*{1mm}%
\hfill%
\vbox{
    \halign{#\hfil        \cr
           IPM/P-2009/004 \cr
           0902.2869\cr
           }
      }
\vspace*{15mm}%

\centerline{{\Large {\bf $3d$ CFT and Multi M2-brane  Theory on $R\by S^2$}}}%
\vspace*{5mm}
\begin{center}
{\bf M. Ali-Akbari}%
\vspace*{0.4cm}

{\it {Institute for Studies in Theoretical Physics and Mathematics (IPM)\\
P.O.Box 19395-5531, Tehran, IRAN}}\\
{E-mails: {\tt aliakbari@theory.ipm.ac.ir}}%
\vspace*{1.5cm}
\end{center}

\begin{center}{\bf Abstract}\end{center}
\begin{quote}
The radial quantization of ${\cal{N}}=8$ theory in three
dimensions is considered i.e. we study the ${\cal{N}}=8$ BLG
theory on $R\times S^2$. We present the explicit from of the
Lagrangian and the corresponding supersymmetry transformations and
supersymmetry algebra. We study spectrum of this theory and some
of its BPS configurations.

\end{quote}%
\end{titlepage}
%
%--------------------------------------------------------------------
%
\section{Introduction}%
The world volume theory of multiple M2-branes has been an open
question over the last two-three decades ago. The low energy limit
of multiple M2-branes theory is expected to be an interacting 2+1
dimensional superconformal field theory with eight transverse
scalar fields as its bosonic content\cite{Schwarz1}. Moreover the
multiple M2-branes theory  should be maximally supersummetric,
which in three dimensions means that it is "${\cal{N}}=8$
supersymmetric theory" and therefore superconformal symmetry group
is $OSp(8|4)$. From AdS/CFT point of view these theories are dual
to $AdS_4\times S^7$ solution of M-theory. The bosonic part of
superalgebra is $SO(8)\times SO(3,2)$ as the global symmetry of
both theories.

Very interesting theory with appropriate symmetries of 3$d$
${\cal{N}}=8$ was proposed by Bagger and Lambert\cite{BL2,BL1,BL3}
and also Gustavsson \cite{Gustavsson}. Therefore this model has
potential to describe world volume of multiple M2-branes. In this
construction the field content is a collection of eight scalars,
fermions and non-propagating gauge fields which are transforming
under 3-algebra and a 4-index structure constant. 3-algebra and
structure constants can be considered as a generalization of a Lie
algebra with triple bracket and 3-index structure constant. The
structure constants satisfy a fundamental identity replacing the
Jacobi identity of a Lie algebra. There should also be a symmetric
invertible metric $h^{ab}$ that can be used to raise and lower
indices. Different aspects of this theory are studied in the
literature\cite{anybody}. Matrix realization of this theory is
also presented in \cite{Gomis,verlinde,AliAkbari,SheikhJabbari}.

According to AdS/CFT and holographic principle this model lives on
the boundary of $AdS_4\times S^7$. It is well known that this
boundary is $R\times S^2$. In this paper we study ${\cal{N}}=8$
BLG model on the $R\times S^2$ background. we construct suitable
supersymmetry transformations and Lagrangian and then investigate
BPS configurations.

This paper is organized as follows. In section 2 we review BLG
model and then in section 3 theory on $R\times S^2$ is considered.
In section 4 we study BPS configurations. Section 5 is devoted to
discussions. The explicit representations for gamma matrices have
been presented in appendix A.
%
%-------------------------------------------------------------------
%
\section{Review of BLG theory}%
To begin, we briefly review Bagger-Lambert construction as a three
dimensional superconformal field theory with $OSp(8|4)$
superalgebra. $spin(8)$ R-symmetry and $spin(4)\equiv spin(3,2)$
conformal symmetry are the bosonic part of superalgebra. Bosonic
fields are $X^I_a$ and non-propagating $A_\mu^{ab}$($\mu=0,1,2$ is
world volume index) and fermionic fields are $\Psi^a$. The index
$I$ labels components of the fundamental $\mathbf{8}_v$
representation of $spin(8)$ as scalar fields corresponding to the
eight directions transverse to M2-branes and $a$ indices take the
values $1,.., dim_{\A}$ with $dim_{\A}$ being the dimension of
3-algebra $\A$ which is yet to be  specified. Representation of
the fermionic fields are $\mathbf{8}_{s}$ and they have two
different indices related to $spin(8)\times spin(3,2)$ which are
suppressed here.

In order to write Lagrangian 4-index structure constants
$f^{abcd}$ is defined associated with a formal, totally
antisymmetric three bracket over 3-algebra generators
\begin{equation}
 [T^a,T^b,T^c] = f^{abc}{}_{d}T^d,
\end{equation}
and inner product is defined by a generalization of the trace over the 3-algebra indices %
\be %
 h^{ab}=Tr(T^aT^b).
\ee %
The 4-index structure constants satisfy the "fundamental identity"
\begin{equation}
 f^{efg}{}_{d}f^{abc}{}_{g}=f^{efa}{}_{g}f^{bcg}{}_{d}
 +f^{efb}{}_{g}f^{cag}{}_{d}+f^{efc}{}_{g}f^{abg}{}_{d}.
\end{equation}
The above bracket and trace satisfy the  %
\be %
 Tr\left([T^a,T^b,T^c]T^d\right)=-Tr\left([T^d,T^a,T^b]T^c\right),
\ee %
implying %
\be\label{antisymmetric} %
f^{abcd}=f^{[abcd]},
\ee %
where $f^{abcd}=f^{abc}_eh^{ed}$. The BLG Lagrangian%
\bea%
 \nonumber {\cal L} &=& -\frac{1}{2}(D_\mu X^{aI})(D^\mu
 X^{I}_{a}) +\frac{i}{2}\bar\Psi^a\gamma^\mu D_\mu \Psi_a
 +\frac{i}{4}\bar\Psi_b\Gamma^{IJ}X^I_cX^J_d\Psi_a f^{abcd}\\
 && - V+\frac{1}{2}\varepsilon^{\mu\nu\lambda}(f^{abcd}A_{\mu
 ab}\partial_\nu A_{\lambda cd} +\frac{2}{3}f^{cda}{}_gf^{efgb}
 A_{\mu ab}A_{\nu cd}A_{\lambda ef}),
\eea%
where %
\bea%
 \nonumber &V&= \frac{1}{12}f^{abcd}f^{efg}{}_d X^I_aX^J_bX^K_cX^I_eX^J_fX^K_g\\
 &(D_\mu X)_a& = \partial_\mu X_a - f^{cdb}{}_{a}A_{\mu\ cd}X_b\equiv\partial_\mu X_a - \tilde A_{\mu}{}^b{}_a X_b,
\eea %
is invariant under the gauge transformations%
\bea\label{gaugetransformation}%
 \nonumber\delta X_a &=& \Lambda_{cd}f^{cdb}{}_{a}X_b \equiv \tilde\Lambda^b{}_{a}X_b\\
 \nonumber\delta \Psi_a &=& \Lambda_{cd}f^{cdb}{}_{a}\Psi_b\\
 \delta(f^{cdb}_{\ \ \ a}A_{\mu cd})&\equiv& \delta\tilde{A}_{\mu}{}^b{}_a=f^{cdb}_{\ \ \
 a}D_\mu\Lambda_{cd},
\eea %
and the supersymmetry variations  %
\bse\label{SUSYtransformationORGINAL}\begin{align}%
 \delta X^I_a&=i\bar{\epsilon}\Gamma^I\Psi_a \\
 \label{SUSYtransformationORGINALpsi}\delta\Psi_a&=D_\mu X^I_a\gamma^\mu\Gamma^I\epsilon-\frac{1}{6}X^I_bX^J_cX^K_df^{bcd}_{\ \ \
 a}\Gamma^{IJK}\epsilon\\
 \delta\tilde{A}^{\ b}_{\mu\ a}&=i\bar{\epsilon}\gamma_\mu\Gamma^IX^I_c\Psi_df^{cdb}_{\ \ \
 a}.
\end{align}\ese%
In the above $\Psi$ and $\epsilon$ should have different $3d$
chirality i.e. $\gamma^{012}\Psi=-\Psi$ and
$\gamma^{012}\epsilon=\epsilon$. It was shown in \cite{BL1,BL3}
that the above supersymmetry transformations are closed up to a
gauge transformation %
\bse\label{9}\begin{align}%
 [\delta_1,\delta_2]X^I_a&=v^\mu\partial_\mu X^I_a+(\tilde{\Lambda}^{b}_{\ a}-v^\nu\tilde{A}_{\nu\ a}^{\ b})X^I_b \\
 [\delta_1,\delta_2]\Psi_a&=v^\mu\partial_\mu\Psi_a+(\tilde{\Lambda}^{b}_{\ a}-v^\nu\tilde{A}_{\nu\ a}^{\ b})\Psi_b \\
 \label{93}[\delta_1,\delta_2]\tilde{A}^{\ b}_{\mu\ a}&=v^\nu\partial_\nu \tilde{A}^{\ b}_{\mu\ a}
 +D_\mu(\tilde{\Lambda}^{b}_{\ a}-v^\nu\tilde{A}_{\nu\ a}^{\ b}),
\end{align}\ese%
where%
\be%
 v^\mu = -2i\bar\epsilon_2\gamma^\mu\epsilon_1,\qquad %
 \tilde\Lambda^b{}_a =-i\bar\epsilon_2\Gamma^{JK}\epsilon_1X^J_c X^K_d f^{cdb}{}_{a}.%
\ee%
It is important to notice that the fundamental identity is
essential to ensure the gauge invariance of the action as well as
the closure of supersymmetry transformations \eqref{93}. Note also
that
the supersymmetry transformations \eqref{9} are written on-shell with the following equations of motion%
\bea%
 \nonumber\gamma^\mu D_\mu\Psi_a+\frac{1}{2}\Gamma^{IJ}X^I_cX^J_d\Psi_bf^{cdb}{}_{a}&=&0\\%
 D^2X^I_a-\frac{i}{2}\bar\Psi_c\Gamma^{IJ}X^J_d\Psi_b f^{cdb}{}_a-\frac{\partial V}{\partial X^{Ia}}&=& 0 \\%
 \nonumber\tilde F_{\mu\nu}{}^b{}_a+\varepsilon_{\mu\nu\lambda}(X^J_cD^\lambda
 X^J_d +\frac{i}{2}\bar\Psi_c\gamma^\lambda\Psi_d )f^{cdb}{}_{a}&=&
 0,
\eea%
that%
\bea%
\tilde F_{\mu\nu}{}^b{}_a  &=&\partial_\nu \tilde A_{\mu}{}^b{}_a
 - \partial_\mu \tilde A_{\nu}{}^b{}_a-\tilde A_{\mu}{}^b{}_c\tilde
 A_{\nu}{}^c{}_a + \tilde A_{\nu}{}^b{}_c \tilde A_{\mu}{}^c{}_a.
\eea%

%
%----------------------------------------------------------------------
%
\section{BLG construction on $R\by S^2$}

To construct the BLG theory on $R\times S^2$, we follow the same
procedure as in \cite{BL1,BL3}. We propose appropriate
supersymmetry transformations and check their closure. As we will
see this fixes all the freedom in the choice of the coefficients
in the supersymmetry variations as well as the equations of
motion. For the "appropriate " supersymmetry variations we need to
work with spinors on $R\times S^2$ which in its own turn is
constructed using the $AdS_4$ fermions. As a result we will show
that supersymmetry closure again demands fundamental identity and
as expected the equation of motion for the $X^I_a$ acquires a mass
term.

\subsection{Killing spinor on $R\by S^2$}

Killing spinor equation on $R\by S^2$ is our aim in this
subsection. The relation between Killing spinor on $AdS_5$ and
$R\times S^3$ has been considered in \cite{RS^3}. Here we follow
the same way to find Killing spinor on $R\times S^2$. In the
global coordinate
the metric of $AdS_4$ with radius $a$ takes the form %
\be %
 ds^2=a^2(-\cosh^2\rho dt^2+d\rho^2+\sinh^2\rho d\Omega_2^2),
\ee %
and Killing spinors defined by %
\be %
\tilde{\nabla}_{\bar{\mu}}\epsilon=\big(\nabla_{\bar{\mu}}-\frac{1}{2R}\gamma_{\bar{\mu}}\big)\epsilon=0. %
\ee %
$\bar{\mu}(=t,\rho,i)$ labels components of $AdS_4$ metric where
$i$ denotes the direction of $S^2$. Supersymmetry parameters are
chiral $4d$ fermions i.e.
$\gamma^{\hat{0}\hat{1}\hat{2}\hat{3}}\epsilon=-\epsilon$ which
have \emph{four complex}  \footnote{Note that after using
$\gamma^{\hat{0}\hat{1}\hat{2}\hat{3}}\epsilon=-\epsilon$,
supersymmetry parameters have \emph{two complex} fermionic degrees
of freedom. By applying
$\gamma^{\hat{0}\hat{1}\hat{2}}\epsilon=\epsilon$ (see after
\eqref{lambdaIJ}), the degrees of freedom are \emph{two real}. }.
Covariant derivative is defined by
$\nabla_{\bar{\mu}}=\partial_{\bar{\mu}}-\frac{1}{4R}\Omega_{\bar{\mu}}^{\hat{a}\hat{b}}\gamma_{\hat{a}\hat{b}}$
that $R$ is radius of 2-sphere and $\Omega^{\hat{a}\hat{b}}$ is
the connection 1-form defined by
$d\omega^{\hat{a}}+\Omega^{\hat{a}}{}_{\hat{b}}\wedge\omega^{\hat{b}}=0$
and $\omega^{\hat{a}}$ is the vierbein defined in the usual manner%
\be %
 g_{\bar{\mu}\bar{\nu}}=\eta_{\hat{a}\hat{b}}\omega^{\hat{a}}_{\bar{\mu}}\omega^{\hat{b}}_{\bar{\nu}},\
 \ \{\gamma^{\bar{\mu}},\gamma^{\bar{\nu}}\}=2g^{\bar{\mu}\bar{\nu}},\ \
 \{\gamma^{\hat{a}},\gamma^{\hat{b}}\}=2\eta^{\hat{a}\hat{b}},
\ee %
and $\tilde{\nabla}$ is written as %
\be\begin{split} %
 \tilde{\nabla}_t&=\partial_t+\frac{1}{2R}\sinh\rho\gamma_t\gamma_\rho-\frac{1}{2R}\cosh\rho\gamma_t=
 e^{-\frac{1}{2R}\rho\gamma}\bigg(\partial_t-\frac{1}{2R}\gamma_t\bigg)e^{\frac{1}{2R}\rho\gamma}\cr
 \tilde{\nabla}_i&=\nabla_i+\frac{1}{2R}\cosh\rho\gamma_i\gamma_\rho-\frac{1}{2R}\sinh\rho\gamma_i=
 e^{-\frac{1}{2R}\rho\gamma}\bigg(\nabla_i-\frac{1}{2R}\gamma_i\gamma\bigg)e^{\frac{1}{2R}\rho\gamma}\cr
 \tilde{\nabla}_\rho&=\partial_\rho-\frac{1}{2R}\gamma_\rho=\partial_\rho+\frac{1}{2R}\gamma.
\end{split}\ee %
We have identified
$\gamma^\rho=-\gamma^{\hat{0}\hat{1}\hat{2}}\equiv-\gamma$,
$\gamma$ is the three dimensional chirality and $\gamma^2=1$. By
above identification, three gamma matrices are independent
describing gamma matrices on $R\by S^2$. Note that in this setup
$SO(8)$ symmetry of the original BLG theory doesn't change.
Therefore, the Killing spinor on $AdS_4$ and Killing spinor on
$R\times S^2$ are related by
\be %
 \epsilon_{AdS_4}=e^{-\frac{1}{2R}\rho\gamma}\epsilon_{R\times
 S^2},
\ee %
where $\epsilon_{R\times S^2}$ satisfies %
\be\label{epsilon-onS2}%
 \nabla_\mu \epsilon=\frac{1}{2R} \omega_\mu \epsilon,  %
\ee%
with%
\be\begin{split}%
 \omega_\mu=(\gamma_t, \gamma_i\gamma)\ &,\ i=1,2 \cr %
 \gamma^\nu\nabla_\nu(\gamma^\mu\nabla_\mu\epsilon)&=-\frac{1}{4}d(d-2)\epsilon,\
 \ d=3
\end{split}\ee%
where $\gamma_i$ are matrices on the $S^2$ and $d$ is space-time
dimension.

\subsection{BLG theory on $R\times S^2$}
Inspired by the BLG and similar analysis for the ${\cal{N}}=4$ on
$R\times S^3$ \cite{RS^3}, we propose the following deformed
supersymmetry transformations for
the ${\cal{N}}=8$ theory on $R\times S^2$ %
\bse\label{SUSYtransformation}\begin{align}%
 \label{SUSYtransformationX}\delta X^I_a&=i\bar{\epsilon}\Gamma^I\Psi_a \\
 \label{SUSYtransformationpsi}\delta\Psi_a&=D_\mu X^I_a\gamma^\mu\Gamma^I\epsilon-\frac{1}{6}X^I_bX^J_cX^K_df^{bcd}_{\ \ \ a}\Gamma^{IJK}\epsilon
 +m\ \Gamma^IX^I_a \gamma^\mu\nabla_\mu \epsilon\\
 \delta\tilde{A}^{\ b}_{\mu\ a}&=i\bar{\epsilon}\gamma_\mu\Gamma^IX^I_c\Psi_df^{cdb}_{\ \ \ a},
\end{align}\ese%
where now $D_\mu$ is covariant derivative on the $R\times S^2$ including gauge field  %
\be%
 (D_\mu X)_a = \nabla_\mu X_a - \tilde A_{\mu}{}^b{}_a X_b,
\ee %
and $m$ is the dimensionless parameter to be fixed later. Instead
of the $3d$ Majorana fermions used in the original BLG analysis
the fermionic fields $\Psi$ should be appropriately chosen for the
$R\times S^2$ case. We choose $\Psi$ to be a chiral fermion on the
$S^2$ and hence $\Psi$ is a one component complex fermion while
also in $\mathbf{8}_s$ of $SO(8)$. For the \susy\ transformation
parameter $\epsilon$ is similarly taken to be a chiral Killing
spinor on $R\times S^2$.%

Closure of the scalar field $X^I$ leads to %
\be\label{closureofX}%
 [\delta_1,\delta_2]X^I_a=v^\mu\partial_\mu X^I_a+(\tilde{\Lambda}^{b}_{\ a}-v^\nu\tilde{A}_{\nu\ a}^{\ b})X^I_b
 +i\Lambda^{IJ}X^J_a,
\ee%
where%
\be\label{lambdaIJ}%
 \Lambda^{IJ}=m\left(\bar{\epsilon}_2\Gamma^{IJ}\gamma^\mu\nabla_\mu\epsilon_1-
\bar{\epsilon}_1\Gamma^{IJ}\gamma^\mu\nabla_\mu\epsilon_2\right)\ .
\ee%
In the above $\gamma^{\hat{0}\hat{1}\hat{2}}\Psi=-\Psi$ and
$\gamma^{\hat{0}\hat{1}\hat{2}}\epsilon=\epsilon$. The
$\Gamma^{IJ}$ term shows the $SO(8)$ R-symmetry rotation
\footnote{Using antisymmetric property of $\gamma^0$,$\gamma^1$
and $\bar{\epsilon}_2\gamma^2\epsilon_1=0$, one can explicitly
show that $\nabla_\mu\Lambda^{IJ}=0$. It means that the R-symmetry
is rigid. Moreover the explicit superalgebra is
written in \eqref{a11} and \eqref{a22} in terms of the fields and their momenta.}. Closure of \susy\ over the fermionic fields leads to%
\be\label{closureofpsi} %
 [\delta_1,\delta_2]\Psi_a=v^\mu\nabla_\mu\Psi_a+(\tilde{\Lambda}^{b}_{\ a}-v^\nu\tilde{A}_{\nu\ a}^{\ b})\Psi_b
 +\frac{i}{4}\Lambda^{IJ}\Gamma^{IJ}\Psi_a\ ,
\ee %
provided that the fermionic equations of motion are%
\be %
 \gamma^\mu
 D_\mu\Psi_a+\frac{1}{2}\Gamma^{IJ}X^I_cX^J_d\Psi_bf^{cdb}{}_{a}=0
 \ ,
\ee %
and that $m=-\frac13$. In other words, the \susy\ closure
condition fixes the only free parameter in our model.

As the last closure condition we examine
$[\delta_1,\delta_2]\tilde{A}^{\ b}_{\mu\ a}$. Upon employing the
fundamental identity,
\be%
  [\delta_1,\delta_2]\tilde{A}^{\ b}_{\mu\ a}=v^\nu\nabla_\nu \tilde{A}^{\ b}_{\mu\ a}
 +D_\mu(\tilde{\Lambda}^{b}_{\ a}-v^\nu\tilde{A}_{\nu\ a}^{\ b}),%
\ee%
provided that $A_\mu$ is satisfying the following equation of motion
\be %
 \tilde F_{\mu\nu}{}^b{}_a+\varepsilon_{\mu\nu\lambda}(X^J_cD^\lambda
 X^J_d +\frac{i}{2}\bar\Psi_c\gamma^\lambda\Psi_d )f^{cdb}{}_{a}=
 0.
\ee %
The above closure conditions establish the supersymmetric
invariance of the BLG model on $R\times S^2$ with the above
modified \susy\ transformations. Note that in this case the \susy\
algebra besides the ``translations on $R\times S^2$'' (the
$\gamma^\mu\nabla_\mu$ term) also involves an $SO(8)$ R-symmetry
rotation.

To find bosonic equation of motion, we take the supervariation of
the fermion equation of motion. This gives
\bea%
  D^2X^I_a-\frac{i}{2}\bar\Psi_c\Gamma^{IJ}X^J_d\Psi_b f^{cdb}{}_a}
  -\frac{1}{4R^2}X^I_a-\frac{\partial V}{\partial X^{Ia}&=& 0. %
\eea%
Finally we present an action for this system. The equations of
motion can be obtained from the action%
\bea%
 \nonumber S&=&\int dtd\Omega_2\sqrt{-g}\Bigg(-\frac{1}{2}(D_\mu X^{aI})(D^\mu
 X^{I}_{a}) +\frac{i}{2}\bar\Psi^a\gamma^\mu D_\mu \Psi_a
 +\frac{i}{4}\bar\Psi_b\Gamma^{IJ}X^I_cX^J_d\Psi_a f^{abcd}\\
 &&-\frac{1}{8R^2}(X^I_a)^2-V+\frac{1}{2}\varepsilon^{\mu\nu\lambda}(f^{abcd}A_{\mu
 ab}\partial_\nu A_{\lambda cd} +\frac{2}{3}f^{cda}{}_gf^{efgb}
 A_{\mu ab}A_{\nu cd}A_{\lambda ef})\Bigg),
\eea%
that $d\Omega_2=R^2\sin\theta d\theta d\phi$. It is not hard to
check that the action is gauge invariant and supersymmetric under
the transformations \eqref{SUSYtransformation}.

In original BLG theory, since $h^{ab}$ is positive definite, it
was proved in \cite{Papadopoulos} that the theory has unique solution which is %
\be %
 f^{abcd}=\epsilon^{abcd},
\ee %
and then the theory has been written as an ordinary gauge theory
with gauge group as $SU(2)\times SU(2)$ with bifundamental matter
\cite{VanRaamsdonk}. It is evident that in our case the theory has
the same structure compared to original theory and therefore it
can be simply written as an ordinary gauge theory with the same
gauge group. Moreover one expects that the $3d\ {\cal{N}}=8$
theory is invariant under the $3d$ parity transformations
$x_0,x_1\rightarrow x_0,x_1$ and $x_2\rightarrow -x_2$ in flat
space. It was shown\cite{BL1} that the parity invariance of
the twisted Chern-Simon term implies that under parity %
\be %
 A_0,A_1\rightarrow A_0,A_1,\ A_2\rightarrow -A_2,\ f\rightarrow
 -f.
\ee %
Parity invariance of the kinetic terms as well as the interaction
terms imply that under parity scalar fields are
invariant and for $3d$ fermions %
\be %
 \Psi^a\rightarrow \gamma^2\Psi^a.
\ee %
By exchanging $(x_0,x_1,x_2)\rightarrow(t,\phi,\theta)$, parity
transformations for the theory on $R\times S^2$ are
$t,\phi,\theta\rightarrow
t,\pi-\phi,\theta$. For the gauge and fermionic fields we have %
\be\begin{split} %
 A_t,A_\theta&\rightarrow A_t,A_\theta,\ A_\phi\rightarrow -A_\phi,\ f\rightarrow -f\cr
 \Psi^a&\rightarrow \gamma^\phi\Psi^a\cr
 X^I_a&\rightarrow X^I_a.
\end{split}\ee %

The original BLG theory enjoys superconformal symmetry. It was
shown that superconformal symmetry can be found by replacing
$\epsilon$ by $\gamma.x\eta$ and adding an appropriate term i.e.
$-X^I\Gamma^I\eta$ to $\delta\Psi_a$\cite{schwarz}. Therefore from
\eqref{SUSYtransformationORGINALpsi} supersymmetry transformation
is
\be\begin{split}\label{susyxi} %
 \delta_\xi\Psi_a\equiv\delta_{susy}\Psi_a=D_\mu
 X^I_a\gamma^\mu\Gamma^I\xi-\frac{1}{6}X^I_bX^J_cX^K_df^{bcd}_{\
 \ \ a}\Gamma^{IJK}\xi,
\end{split}\ee %
and superconformal transformation is %
\be\begin{split}\label{superconformal} %
 \delta_\eta\Psi_a\equiv\delta_{su.conf.}\Psi_a&=D_\mu
 X^I_a\gamma^\mu\Gamma^I\gamma.x\eta-\frac{1}{6}X^I_bX^J_cX^K_df^{bcd}_{\
 \ \ a}\Gamma^{IJK}\gamma.x\eta
 -\ \Gamma^IX^I_a\eta,
\end{split}\ee %
where $\xi$ and $\eta$ are constant spinors. It is easy to write
supersymmetry transformations on $R\times
S^2$\eqref{SUSYtransformationpsi} in terms of \eqref{susyxi} and
\eqref{superconformal}, as a combination of $3d$ superPoincare
and $3d$ superconformal transformations%
\be\label{superconformalonRS2} %
 \delta_{\epsilon}\Psi_a=\delta_{\xi}\Psi_a+\delta_{\eta}\Psi_a.
\ee %
This leads %
\be\begin{split} %
 \epsilon=\xi+\gamma.x\eta.
\end{split}\ee %
In the original BLG theory $\xi$ and $\eta$ are $3d$ Majorana
fermion in $\mathbf{8}_c$ of $SO(8)$ and then they have $16+16$
degrees of freedom. Supersymmetry transformations on $R\times S^2$
are generated by $16$ independent $\epsilon$'s. The other
combination should be considered as a "\emph{superconformal
symmetry}" on $R\times S^2$.

In order to understand the theory we would like to study complete
spectrum about $X^I=0$ vacuum. To do so, we expand the theory
about the vacua to second order in small
fluctuations. Then equations of motion for $X^I_a$ are %
\be %
 \big(\partial_t^2-\frac{1}{R^2}\nabla_{S^2}^2+\frac{1}{4R^2}\big)X^I_a=0,
\ee %
where $\nabla_{S^2}^2$ is written on the sphere with radius one.
By expanding $X^I_a$ in terms of spherical harmonics on the
2-sphere we have %
\be\label{XYLM} %
 X^I_a=\sum_lx^{I}_{a,lm}e^{iw_lt}Y_{lm}(\theta,\phi),
\ee %
and hence these modes would have mass squared equal to %
\be %
 R^2w_l^2=(l+\frac12)^2,\ \ l=0,1,...
\ee %
For fermionic fields by using \eqref{SUSYtransformationX} we have %
\be %
 \big(\partial_t^2-\frac{1}{R^2}\nabla_{S^2}^2+\frac{1}{4R^2}\big)\delta
 X^I_a=0,
\ee %
which leads\footnote{Note that $\nabla^2\epsilon=-\frac{\epsilon}{4R^2}$.} %
\be %
 \big(\partial^2_t-\frac{1}{R^2}\nabla^2_{S^2}+\frac{1}{4R^2}\big)\Psi_a=0.
\ee %
Making the expansion \cite{harmonic} %
\be %
 \Psi_a=\sum_j\psi^{jm}_ae^{i\omega_l t}(\sin\theta)^{|m|}Y_{jm}(\theta,\phi), %
\ee %
where quantum number $j$ is total angular
momentum($j=l\pm\frac12$)
of fermions. Hence mass squared is %
\be\begin{split} %
 j=l+\frac12\ \ \ \ :\ R^2\omega^2_l&=(l+1)^2,\ \ l=0,1,... \cr%
 j=l-\frac12\ \ \ \ :\ R^2\omega^2_l&=l^2,\ \ \ \ \ \ \ \ \ l=1,2,...
\end{split}\ee %
As a result of supersymmetry the sum of boson masses and the sum
of fermion masses are both $16(l+\frac12)^2$.

Recently, in \cite{Aharony} an infinite class of brane
configurations was given whose low energy effective Lagrangian is
a Chern-Simon theory with $SO(6)$ R-symmetry and ${\cal{N}}=6$
supersymmetry. These theories are related to N M2-branes in
$R^8/Z_k$ including $k=1$. After that by relaxing the condition on
three-bracket so that it is no longer real and antisymmetric in
all three indices i.e. %
\be\label{f} %
 f^{ab\bc\bd}=-f^{ba\bc\bd},\ \ f^{ab\bc\bd}=f^{*\bc\bd ab}.
\ee %
${\cal{N}}=6$ theories based on 3-algebra have been obtained
\cite{Bagger,SheikhJabbari}. However the new three- bracket is
still required to satisfy the fundamental identity. The
supersymmetry transformations are\cite{Bagger}%
\begin{eqnarray}\label{finalsusy}
\nonumber  \delta Z^A_d &=& i\bar\epsilon^{AB}\psi_{Bd} \\
\nonumber  \delta \psi_{Bd} &=& \gamma^\mu D_\mu
Z^A_d\epsilon_{AB} +
  f^{ab\bc}{}_dZ^C_a Z^A_b \bZ_{C\bc} \epsilon_{AB}+
  f^{ab\bc}{}_d Z^C_a Z^D_{b} \bZ_{B\bc}\epsilon_{CD} \\
  \delta \tilde A_\mu{}^c{}_d &=&
-i\bar\epsilon_{AB}\gamma_\mu Z^A_a\psi^B_\bb f^{ca\bb}{}_d +
i\bar\epsilon^{AB}\gamma_\mu \bZ_{A\bb}\psi_{Ba} f^{ca\bb}{}_d ,
\end{eqnarray}
%where $Z^A,\psi_A$ and $\epsilon_{AB}$ are in
%$\textbf{4}$,$\bar{\textbf{4}}$ and $\textbf{6}$ representation of
%$SO(6)$ respectively.
where $\epsilon_{AB}$ is in the $\textbf{6}$ of $SU(4)$ and a
raised $A$ index indicates that the field is in the $\textbf{4}$
of $SU(4)$; a lowered index transforms in the $\bar{\textbf{4}}$.
One can write above theory on $R\times S^2$ by adding an
appropriate mass term, i.e.
$-\frac{1}{3}Z^A_d\gamma^\mu\nabla_\mu\epsilon_{AB}$, in variation
of fermionic fields. Since the antisymmetry condition was not used
in our earlier supersymmetry closure analysis the above
supersymmetry transformations plus mass term will still remain
closed. In particular, the closure of the scalar fields will
exactly work in the same way as in the ${\cal{N}}=8$ theory. For
the closure of gauge fields, equation of motion, \eqref{f} and
fundamental identity are enough. The closure of fermionic fields
just requires fermionic equation of motion. The equation of motion
for scalars $Z^A_a$, as before is found by taking the
supervariation of the fermion equation of motion if we apply
\eqref{f}. Therefore, one can reproduce the ${\cal{N}}=6$
supersymmetric theories on $R\times S^2$. (Since the computations
are essentially the same as the ${\cal{N}}=8$ we do not repeat the
equations.)%

Finally, superalgebra may be written by using \eqref{closureofX}
and \eqref{closureofpsi}. As we explained before fermionic fields
have two different indices relating to $SO(3)\times U(1)$ and
$SO(8)$ which is the bosonic part of $OSp(8|2)\times U(1)$
superalgebra. Let's label them with $\dot{\alpha}=1,2$ and
$\dot{A}=1,..,8$
respectively. Then the superalgebra is %
\be\label{a11} %
 \{Q^{\dot{A}}_{\dot{\alpha}},Q^{\dot{B}}_{\dot{\beta}}\}=-2\delta^{\dot{A}\dot{B}}
 (\gamma^\mu\gamma^0)_{\dot{\alpha}\dot{\beta}}P_\mu+\frac12\delta_{\dot{\alpha}\dot{\beta}}(\Gamma^{IJ})^{\dot{A}\dot{B}}J^{IJ},
\ee %
where %
\be\label{a22}\begin{split} %
 J^{IJ}&=\int d\Omega_2\sqrt{-g}\bigg(X^IP^J-X^JP^I+\frac12\psi^\dagger(i\Gamma^{IJ})\psi\bigg)\cr%
 Q&=\int d\Omega_2\sqrt{-g}\bigg(D_\mu X^I_a\Gamma^I\gamma^\mu\gamma^0\Psi^a-\frac{1}{6}X^I_bX^J_cX^K_df^{bcd}_{\ \ \
 a}\Gamma^{IJK}\gamma^0\Psi_d
 +\frac{1}{4R}\ \Gamma^IX^I_a\gamma^0\Psi^a\bigg).
\end{split}\ee %
We have fixed that $J^{IJ}$'s are $SO(8)$ generators. The
superalgebera for original BLG theory has been discussed in
\cite{superalgebra}.
%
%------------------------------------------------------------------------
%
\section{BPS Solution}
By definition a BPS configuration is a state which is invariant
under some specific supersymmetry transformations. For the
configurations in which spinor fields are turned off the
non-vanishing supersymmetry variations are only
$\delta_\epsilon\Psi_a$ and hence
BPS equations read as %
\be\label{bpsequation1} %
 \delta_\epsilon\Psi_a=0,
\ee %
for arbitrary $\epsilon$. From the above equation and
\eqref{SUSYtransformationpsi} it is clear that the $X^I=0$ vacuum
is a full BPS configuration(with 32 supercharges). Another class
of BPS solutions are small fluctuations about vacuum. In this case
the equation
\eqref{bpsequation1} reads as %
\be %
 \big(\gamma^\mu\nabla_\mu
 X^I_a\Gamma^I-\frac{1}{2R}X^I_a\Gamma^I\big)\epsilon=0,
\ee %
where gauge and fermionic fields are turned off. Replacing from \eqref{XYLM} we have%
\be %
 \bigg(\gamma^t(i\omega_l)+\gamma^i\partial_i-\frac{1}{2R}\bigg)X^I_a\Gamma^I\epsilon=0,\
 \ i=\theta,\phi
\ee %
which evidently is right just for $l=0$ bosonic fluctuations and
then they are $1/4$ BPS configurations. In this case we have a
short multiplet including eight bosonic and four fermionic degrees
of freedom. Other possibilities of $l$ are non-BPS solutions with
equal number of bosonic and fermionic degrees of freedom. Hence
$(l,l+\frac12,l+1),\ l>0$ assemble to a long multiplet. In what
follows we discuss other classes of 1/4 BPS configurations.

\subsection{1/4 BPS configuration}
Let us start with the case in which $X^{5,6,7,8}$'s are turned
off and then BPS equation \eqref{bpsequation1} takes the form %
\be\label{bpsequation} %
 \big(\gamma^\mu D_\mu X^{\hat{i}}\Gamma^{\hat{i}}-\frac{1}{6}[X^{\hat{i}},X^{\hat{j}},X^{\hat{k}}]
 \Gamma^{\hat{i}\hat{j}\hat{k}}-\frac{1}{2R}X^{\hat{i}}\Gamma^{\hat{i}}\big)\epsilon=0,\
 \hat{i}=1,2,3,4.
\ee %
In order to solve above equation we introduce
\be %
 X^{\hat{i}}=\alpha\Gamma^{\hat{i}}.%
\ee %
$\Gamma^{\hat{i}}$'s are in $n\times n$ representation of
$Spin(4)$ and obey %
\be %
 [\Gamma^{\hat{i}},\Gamma^{\hat{j}},\Gamma^{\hat{k}}]=12\epsilon^{\hat{i}\hat{j}\hat{k}\hat{l}}\Gamma^{\hat{l}} ,%
\ee %
and $\alpha$ is a dimensional constant. Therefore, the first term
in \eqref{bpsequation} vanishes and it leads to
\be\label{degree} %
 \big(2.3!\alpha^2\Gamma^5-\frac{1}{2R}\id \big)X^{\hat{i}}\Gamma^{\hat{i}}\epsilon=0, %
\ee %
which has a solution if $\alpha^2=\frac{1}{24R}$\ (\ $\Gamma^5$ is
the SO(4) chirality matrix). These solutions are exactly fuzzy
three sphere with $SO(4)$ symmetry explained in the literature
e.g. \cite{Tiny}. One expects that the theory which lives on the
two membranes can be described by BLG theory. It means that in our
solution membranes blow up to a fuzzy three sphere in transverse
directions. \eqref{degree} shows that $\epsilon$ has eight real
fermionic degrees of freedom and  our solutions are 1/4 BPS. We
reproduce trivial case $X^{I}=0$ when $R$ goes to infinity.
%\subsection{1/2 BPS configuration}

The other case happens when $\alpha$ is not a constant and can
vary on the 2-sphere in the $\theta$ direction. We
then have %
\be %
 \gamma^\theta\partial_\theta X^{\hat{i}}\Gamma^{\hat{i}}-\frac{1}{6}[X^{\hat{i}},X^{\hat{j}},X^{\hat{k}}]\Gamma^{\hat{i}\hat{j}\hat{k}}
 -\frac{1}{2R}X^{\hat{i}}\Gamma^{\hat{i}}=0. %
\ee %
It is straightforward to check that the above equation is solved
with %
\be %
 X^{\hat{i}}=\alpha(\theta)\Gamma^{\hat{i}}, %
\ee %
provided that %
\be\label{2} %
 \alpha(\theta)=\frac{1}{\sqrt{24s_1R(1-e^{s_2(\theta-\theta_0)}})},
\ee %
that we have used %
\bea %
 \Gamma^{5}\epsilon=s_1\epsilon\cr
 \gamma^\theta\epsilon=s_2\epsilon,
\eea %
where $s_1$ and $s_2$ can independently be $+1$ or $-1$. Two
different cases exist here which are
$e^{s_2(\theta-\theta_0)}>1,s_1=-1$ and
$e^{s_2(\theta-\theta_0)}<1,s_1=+1$. Regarding to the sign of
$s_2$ in each case there are eight independent $\epsilon$'s and
therefore these configurations are $1/4$ BPS. These solutions
correspond to M2-brane along $0\theta\phi$ ending on M5-brane
along $01234\phi$ which means that M5-brane wraps in $\phi$
direction and as a result there is a $U(1)_\phi$ symmetry. Unlike
the previous $1/4$ BPS configurations these family of solutions
change to Basu-Harvey configurations \cite{Basu} in specific limit
as an open membrane ending on M5-brane(see also \cite{Krishnan}).
The "Basu-Harvey limit" is then a limit where $R$ is taken to
infinity, keeping $x$ finite,
i.e. %
\be %
  R\rightarrow\infty,\ \ \ \theta=x/R,x\ finite,
\ee %
and \eqref{2} becomes %
\be %
  \alpha(x)=\frac{1}{\sqrt{-24s_1s_2(x-x_0)}}.
\ee %
If $x>x_0$ then we should take $s_1s_2=-1$ indicating
$s_1=+1,s_2=-1$ or $s_1=-1,s_2=+1$. Each of them preserves four
independent $\epsilon$'s and we have 1/4 BPS Basu-Harvey
configurations. (For the other case, $x<x_0$, there are again
eight $\epsilon$'s.)
%
%------------------------------------------------------------------------
%
\section{Conclusion}
In this work we have generalized the $3d,\ {\cal{N}}=8$ BLG theory
on flat space to $R\times S^2$. As we discussed an additional term
adds to supersymmetry transformation of fermion and also
supersymmetry parameters are no longer constant and vary on the
2-sphere. These two differences have two results. The first one
appears in the closure of bosonic and fermionic fields that we
have a $SO(8)$ R-symmetry rotation. This rotation didn't appear
for gauge fields because they have singlet representation of
$SO(8)$. Appearing a new term in the equation of motion for $X$'s
leaded a mass term in the Lagrangian is the second one. However
the equations of motion for gauge and fermionic fields formally
remain unchange. Our theory like original BLG theory is parity
invariance as expected. We have also considered small fluctuation
about vacuum and superalgebra .

It was argued that ABJM model can be written on $R\times S^2$.
Although $f^{abcd}$ is not real and fully antisymmetric the
supersymmetry transformation including mass term closes up to a
gauge transformation.

In the last section we have studied BPS configurations. One family
of $1/4$ BPS configurations are fuzzy three sphere with $SO(4)$
symmetry and another one can be considered as M5-M2 configuration
which M5 has been wrapped in the $\phi$ direction. In the
Basu-Harvey limit this family of solutions reproduce Basu-Harvey
configuration as an open membrane ending on M5-brane.

\subsection*{Acknowledgment}
It is a great pleasure to thank M.M. Sheikh-Jabbari for several
insightful discussions and useful comments.

\appendix
\section{Gamma matrices}
In this appendix we briefly consider our notation of
$\Gamma$-matrices. The eleven dimensional $\Gamma$-matrices are
defined by %
\be %
 \{\Gamma^M,\Gamma^N\}=2\eta^{MN},\ \ M,N=0,..,10 %
\ee %
where $\eta^{MN}=diag(-,+^{10})$. Under dimension reduction to
three dimensions we have %
\be %
 SO(10,1)\supset SO(2,1)\times SO(8) %
\ee %
\be\begin{split} %
 \{\gamma^\mu,\gamma^\nu\}&=2\eta^{\mu\nu},\ \ \mu,\nu=0,1,2\cr%
 \{\Gamma^I,\Gamma^J\}&=2\delta^{IJ},\ \ I,J=1,..,8 \cr %
 \{\Gamma^I,\gamma^\mu\}&=0
\end{split}\ee
\be\begin{split} %
 \gamma^\mu&=\bar{\gamma}^\mu\otimes\gamma^8\cr%
 \Gamma^I&=\id_4\otimes\gamma^I
\end{split}\ee %
where %
\be\begin{split} %
 \bar{\gamma}^\mu=\left(%
 \begin{array}{cc}
  0 & i\tau^\mu \\
  -i\tau^\mu & 0 \\
 \end{array}%
 \right),\ \ \tau^0&=i\sigma^3,\tau^1=\sigma^1,\tau^2=\sigma^2 \cr%
 \{\gamma^I,\gamma^8\}&=0,\ \ (\gamma^8)^2=1
\end{split}\ee %

\end{document}